\documentclass[a4paper, twocolumn]{article}
\usepackage[utf8]{inputenc}
\usepackage{cite}
\usepackage{graphicx}
\usepackage{amssymb}
\usepackage{amsfonts}
\usepackage{amsmath}
\usepackage{bm}
\usepackage{longtable}
\usepackage{epsfig}
\usepackage{multirow}
\usepackage{authblk}%
\usepackage[paperwidth=19.5cm,paperheight=27.0cm,top=2.0cm,bottom=2.0cm,left=1.5cm,right=1.5cm]{geometry}
\usepackage[switch]{lineno}

\usepackage{fancyhdr}
\begin{document}

\title{Linear-in-temperature resistivity and Planckian dissipation arise in a stochastic quantization model of Cooper pairs }
\author{Xiao-Song Wang}
\setcounter{footnote}{0}
\affil{{\normalsize Institute of Mechanical and Power Engineering, Henan Polytechnic University, Jiaozuo, Henan Province, 454000, China}}%
\date{Mar. 2nd, 2024}

\twocolumn [
	\begin{@twocolumnfalse}
		\maketitle
		\begin{abstract}
			\newgeometry{left=1.0cm, right=1.0cm}%
We suppose that a Cooper pair (CP) will experience a damping force exerted by the condensed matter. A Langevin equation of a CP in two dimensional condensed matter is established. Following a method similar to Nelson's stochastic mechanics, generalized Schr\"{o}dinger equation of a CP in condensed matter is derived. If the CPs move with a constant velocity, then the corresponding direct current (DC) electrical conductivity can be calculated. Therefore, a Drude like formula of resistivity of CPs is derived. We suppose that the damping coefficient of CPs in two dimensional cuprate superconductors is a linear function of temperature. Then the resistivity and scattering rate of CPs turn out to be also linear-in-temperature. The origin of linear-in-temperature resistivity and Planckian dissipation in cuprate superconductors may be the linear temperature dependence of the damping coefficient of CPs.

\

keywords: Planckian resistivity; Planckian dissipation; Cooper pair; strange metal; stochastic mechanics.
	  \end{abstract}
	\end{@twocolumnfalse}

\



\

\

]

\section{Introduction  \label{sec 100}}
\newtheorem{assumption}{\bfseries Assumption}
\newtheorem{definition}[assumption]{\bfseries Definition}
\newtheorem{lemma}[assumption]{\bfseries Lemma}
\newtheorem{proposition}[assumption]{\bfseries Proposition}
\newtheorem{theorem}[assumption]{\bfseries Theorem}
\newtheorem{wcorollary}[assumption]{\bfseries Corollary}

It is known that the resistivity $\rho$ of the normal states of cuprate superconductors obeys the following relationship \cite{BruinJAN2013,VarmaCM2020,TaupinM2022,ZangQ2023}
\begin{equation}\label{resistivity 100-100}
\rho = \rho_{0} + AT,
\end{equation}
where $\rho_{0}$ is the residual resistivity, $A$ is a coefficient independent of temperature, $T$ is temperature.

These phenomena of linear temperature dependence of resistivity are found in numerous strongly correlated electron
systems, such as the heavy fermion compounds \cite{BruinJAN2013,TaupinM2022}, transition metal oxides \cite{BruinJAN2013,AtaeiA2022,ZangQ2023}, iron pnictides \cite{BruinJAN2013}, magic angle twisted bilayer graphene, organic metals \cite{BruinJAN2013} and conventional metals \cite{BruinJAN2013}, often in connection with unconventional superconductivity. Sometimes this linear-in-temperature resistivity is called Planckian resistivity \cite{VarmaCM2020}. When superconductivity is destroyed by a high magnetic field, the recovered normal state still obeys this law of linear-in-temperature resistivity in the low temperature region \cite{ZangQ2023}. In most of the heavy fermion materials, the linear-in-temperature resistivity appears when they have been tuned by some external parameter to create a low-temperature continuous phase transition which is referred to a quantum critical point (QCP) \cite{BruinJAN2013}. Thus, the linear temperature dependence of resistivity are often associated with quantum criticality. The linear-in-temperature resistivity of $LSCO$ with different gradients, different doping dependencies and different origins appears not only at high temperature but also at low temperature \cite{HusseyNE2011}.

Strange metal behavior refers to a linear temperature dependence of the electrical resistivity \cite{BruinJAN2013,TaupinM2022}. A unified theory of this scaling law (\ref{resistivity 100-100}) in different strange metals is still an open problem \cite{VarmaCM2020,TaupinM2022}.

Before the discovery of quantum mechanics, a successful formula of resistivity of metals is proposed in the Drude model (\cite{AshcroftNW1976}, p.\ 7). Shortly after the discovery of quantum mechanics, Sommerfeld improved the Drude model. In the Sommerfeld model, the following Drude formula of resistivity of metals can be derived approximately based on quantum theoty (\cite{AshcroftNW1976}, p.\ 251)
\begin{equation}\label{Drude 100-200}
\rho = \frac{m^{*}}{ne^{2}\tau},
\end{equation}
where $n$ is the number dendity of electrons, $e$ is the electric charge of an electron, $m^{*}$ is the effective mass of an electron, $\tau$ is the relaxation time of an electron.

If the transport scattering rate $1/\tau$ is linear-in-temperature and is the only temperature-dependent quantity in Eq.(\ref{Drude 100-200}), then the scaling law (\ref{resistivity 100-100}) of resistivity can be derived directly. Thus, a clue to study the scaling law (\ref{resistivity 100-100}) is to investigate the relaxation time $\tau$ in the Drude formula (\ref{Drude 100-200}).

The Drude formula (\ref{Drude 100-200}) is valid only for charge carriers which obey the Fermi-Dirac distribution. Experiments have shown that the dominant charge carriers in $YBa_{2}Cu_{3}O_{7-\delta}$ (YBCO) film are Cooper pairs (CPs) \cite{YangC2022}. Since CPs are not Fermions, the Drude formula (\ref{Drude 100-200}) may be not valid in the normal states of cuprate superconductors. Thus, an interesting question is that whether a similar formula for the resistivity of the normal states of cuprate superconductors exists.

According to the Heisenberg uncertainty principle, a local equilibration time of any many-body quantum system cannot be faster than the following Planckian time $\tau_{p}$ \cite{TaupinM2022}
\begin{equation}\label{Planckian 100-300}
\tau_{p} = \frac{\hbar}{k_{B}T},
\end{equation}
where $h$ is the Plank constant, $\hbar = h/2\pi$, $k_{B}$ is the Boltzmann constant.

This timescale $\tau_{p}$ is associated with quantum criticality and known to bound the validity of a Boltzmann description of transport \cite{HartnollSA2022}. $\tau_{p}$ is suggested to be the lower bound of the phase coherence time in quantum critical systems \cite{HartnollSA2022}. $\tau_{p}$ is also known to control the electronic dynamics of the cuprate strange metal \cite{HartnollSA2022}. Thus, an idea is that the relaxation time $\tau$ of CP in cuprate superconductors may be proportional to the Planckian time $\tau_{p}$, i.e., $\tau=\alpha_{0}\tau_{p}$, where $\alpha_{0}$ is a dimensionless parameter. Indeed, experiments have shown that the scattering rate $1/\tau$ in the region of the temperature-linear resistivity of a wide range of metals, including heavy fermion, oxide \cite{AtaeiA2022,YangC2022,ZangQ2023}, pnictide, organic metals and conventional metals, can be written as \cite{BruinJAN2013}
\begin{equation}\label{scatter 100-400}
\frac{1}{\tau} = \frac{\alpha_{0}k_{B}T}{\hbar},
\end{equation}
where $\alpha_{0} \approx 1$.

Eq.(\ref{scatter 100-400}) shows that the relaxation time $\tau$ is approximately equal to the Planckian time $\tau_{p}$, i.e.,  $\tau \approx \tau_{p}$. The case of $\alpha_{0} \approx 1$ is referred to the Planckian dissipation \cite{TaupinM2022}. It is surprising that the linear-in-temperature scattering rate $1/\tau$ and the behaviors of Planckian dissipation in these materials (except the conventional metals) can be seen down to low temperatures with appropriate tuning by magnetic field, chemical composition or hydrostatic pressure \cite{BruinJAN2013}. It is suggested that there may be a fundamental principle governing the transport of CPs \cite{YangC2022}.

If Eq.(\ref{scatter 100-400}) and Eq.(\ref{Drude 100-200}) are valid in the normal states of cuprate superconductors, then Eq.(\ref{resistivity 100-100}) may be derived. In this manuscript we focus on this clue and try to derive the scaling law (\ref{resistivity 100-100}).

\section{Stochastic mechanics of a Cooper pair in two dimensional condensed matter \label{sec 300}}
In order to explain the energy quantization of atoms, E. Schr\"{o}dinger proposes the following equation for a non-relativistic particle moving in a potential \cite{Landau-Lifshitz1958}
\begin{equation}\label{Schrodinger 100-100}
i \hbar \frac{\partial \psi}{\partial t} =
-\frac{\hbar^2}{2m}\nabla^2\psi + U(\mathbf{r})\psi,
\end{equation}
where $t$ is time, $\mathbf{r}$ is a point in space, $\psi(\mathbf{r},t)$ is the wave function, $m$ is the mass of the particle, $U(\mathbf{r})$ is the potential, $h$ is the Plank constant, $\hbar = h/2\pi$, $\nabla^2 \equiv \partial^2/\partial r_{1}^2 +\partial^2/\partial r_{2}^2 + \partial^2/\partial r_{3}^2$ is the Laplace operator in a Cartesian coordinate $\{r_{1}, r_{2}, r_{3}\}$.

The Schr\"{o}dinger equation (\ref{Schrodinger 100-100}) is a fundamental assumption in non-relativistic quantum mechanics \cite{Landau-Lifshitz1958}. Although the Schr\"{o}dinger equation can be used to describe some non-relativistic quantum phenomena, the origin of quantum phenomena remains an unsolved problem in physics for more than 100 years \cite{Schilpp1949,Jammer1974}. Although the axiomatic system of quantum mechanics was firmly established, the interpretation of quantum mechanics is still open \cite{Schilpp1949,Jammer1974}. There exist some paradoxes in quantum mechanics \cite{Einstein1935,Bohr1935,Tarozzi-Merwe1988,Selleri1990}, for instance, the paradox of reduction of a wave packet and the paradox of the Schr\"{o}dinger cat.

F\'{e}nyes proposed an interpretation of quantum mechanics based on a Markov process. F\'{e}nyes' work was developed by Weizel and discussed by Kershaw \cite{Kershaw1964}. According to Luis de Broglie \cite{Broglie1956}, the success of the probabilistic interpretation of $|\psi|^2$ inspired Einstein to speculate that the probability $|\psi|^2$ is generated by a kind of hidden Brownian motions of particles. This kind of hidden motions was called quasi-Brownian motions by Luis de Broglie \cite{Broglie1956}.

If the quantum phenomena stem from the stochastic motions of particles, then we may establish a more fundamental and more powerful theory of quantum phenomena other than quantum mechanics. The Schr\"{o}dinger equation may no longer be a basic assumption and may be derived in this new theory. Indeed, E. Nelson \cite{Nelson1966} derived the Schr\"{o}dinger equation by means of theory of stochastic processes based on the assumption that every particle with mass $m$ in vacuum is subject to Brownian motion with diffusion constant $\hbar / 2m$.

Inspired by Nelson's stochastic mechanics \cite{Nelson1966,Nelson1972,Nelson1985,Guerra-Ruggiero1973,Guerra1981,Guerra-Morato1983,Guerra-Marra1983,Guerra-Marra1984,Namsrai1986,Guerra1995}, we propose a theoretical derivation of the Schr\"{o}dinger equation based on Newton's second law and a mechanical model of vacuum \cite{WangXS2014}.

Recently, monolayer crystals of the high-temperature superconductor $Bi_{2}Sr_{2}CaCu_{2}O_{8+\delta}$ (Bi-2212) was obtained by a fabrication process \cite{YuY2019}. The superconductivity, the pseudogap, charge order and the Mott state at various doping concentrations of the monolayer Bi-2212 reveals that the phases are indistinguishable from those in the bulk \cite{YuY2019}. Monolayer Bi-2212 displays the fundamental physics of cuprate superconductors \cite{YuY2019}. Therefore, monolayer copper oxides is a platform for studying high-temperature superconductivity in two dimensions. Thus, we focus on two dimensional condensed matters.

Modern experiments, for instance, the Casimir effect \cite{CasimirH1948,SparnaayMJ1957}, have shown that vacuum is not empty. Thus, we suppose that there is a damping force exerted on each particle by vacuum \cite{WangXS2014}. For a microscopic particle moving in vacuum, we have the following relation \cite{WangXS2014}
\begin{equation}\label{planck 50-60}
\hbar = \frac{2 k_{0}T_{0}}{\eta_{0}},
\end{equation}
where $k_{0}$ is a constant similar to the Boltzmann constant $k_{B}$, $T_{0}$ is the temperature of the $\Omega(0)$ substratum in the vicinity of the particle in vacuum \cite{WangXS200810}, $\eta_{0}$ is a damping coefficient related to vacuum.

It is known that a CP in condensed matter may be scattered by ions, electrons, phonon, etc. In the Drude theory of metals, the effect of individual electron collisions is approximately treated by introducing a damping force into the equation of motion of an electron (\cite{AshcroftNW1976}, p.\ 11). Following the Drude theory, we suppose that a CP in a condensed matter will experiences not only a damping force exerted by vacuum but also an additional damping force exerted by the condensed matter. We introduce a two dimensional Cartesian coordinate system $\{r_{1}, r_{2}\}$ which is attached to the condensed matter. We suppose that the two dimensional velocity $\mathbf{v}=d\mathbf{r}/dt$ of the CP exists. Applying Newton's second law, the motion of a CP may be described by the following Langevin equation \cite{Chandrasekhar1943}
\begin{eqnarray}
m_{c}\frac{d^2\mathbf{r}}{dt^2} &=& - \eta_{0}m_{c}\mathbf{v} - \eta_{1}m_{c}\mathbf{v} \nonumber \\
 &&- \eta_{2}m_{c}\frac{d^2\mathbf{r}}{dt^2} +\mathbf{F}(\mathbf{r},t) +\mathbf{\xi}(t),\label{Langevin 300-100}
\end{eqnarray}
where $m_{c}$ is the mass of the CP, $\eta_{1}$ is a damping coefficient related to the condensed matter, $\eta_{2}$ is a quasi-inertial force coefficient, $\mathbf{\xi}(t)$ is a two dimensional random force and $\mathbf{F}(\mathbf{r},t)$ is a two dimensional external force field.

We introduce the following definitions
\begin{eqnarray}
- \eta_{0}m_{c}\mathbf{v} - \eta_{1}m_{c}\mathbf{v} &=& - \eta_{0}m_{d}\mathbf{v}, \label{damping 300-200}\\
m_{c}\frac{d^2\mathbf{r}}{dt^2} + \eta_{2}m_{c}\frac{d^2\mathbf{r}}{dt^2} &=& m_{q}\frac{d^2\mathbf{r}}{dt^2}, \label{inertial 300-300}
\end{eqnarray}
where $m_{d}$ is the damping mass of the CP, $m_{q}$ is the quasi-inertial mass of the CP.

Eq.(\ref{damping 300-200}) and Eq.(\ref{inertial 300-300}) can be written as
\begin{eqnarray}
m_{d} &=& \frac{\eta_{0}+\eta_{1}}{\eta_{0}}m_{c}, \label{damping 300-250}\\
m_{q} &=& (1+\eta_{2})m_{c}, \label{inertial 300-350}
\end{eqnarray}

Using Eq.(\ref{damping 300-200}) and Eq.(\ref{inertial 300-300}), Eq.(\ref{Langevin 300-100}) can be written as
\begin{equation}\label{Langevin 300-200}
m_{q}\frac{d^2\mathbf{r}}{dt^2} = - \eta_{0}m_{d}\mathbf{v} +\mathbf{F}(\mathbf{r},t) +\mathbf{\xi}(t).
\end{equation}

Let $\xi_{i}(t)$ be the $i$th component of the random force $\mathbf{\xi}(t)$, i.e., $\mathbf{\xi}(t)=(\xi_{1}(t), \xi_{2}(t))$.
\begin{assumption}\label{assumption 300-100}
Assume that the force field $\mathbf{F}(\mathbf{r},t)$ is a continuous function of $\mathbf{r}$ and $t$. Inspired by the Ornstein-Uhlenbeck theory \cite{UhlenbeckOrnstein1930,Kallenberg1997} of Brownian motion, we suppose that the random force $\mathbf{\xi}$ exerted on the CP by the condensed matter is a two-dimensional Gaussian white noise \cite{Gelfand-Vilenkin1961,Soong1973,Arnold1974,Gardiner2004} and the variance $E(\xi_{i}^{2})$ of the $i$th component of $\mathbf{\xi}$ is \cite{WangXS2014}
\begin{equation}\label{variance 200-300}
E(\xi_{i}^{2}(t)) \equiv \sigma_{i}^{2} = 2 \eta_{0}m_{d} k_{0} T_{\omega},
\end{equation}
where $\sigma_{i}>0$, $i=1, 2$, $k_{0}$ is a parameter similar to the Boltzmann constant, $T_{\omega}$ is the temperature of the $\Omega(0)$ substratum \cite{WangXS200810} in the location of the condensed matter.
\end{assumption}

For convenience, we introduce the following notation
\begin{equation}\label{diffusion 200-400}
D_{1} = \frac{\sigma_{1}^{2}}{2} = \eta_{0} m_{d} k_{0} T_{\omega}.
\end{equation}

The Gaussian white noise $\mathbf{\xi}$ is the generalized derivative of a Wiener process $\mathbf{Q}(t)$ \cite{Gelfand-Vilenkin1961,Arnold1974}. We can write formally \cite{Gelfand-Vilenkin1961,Soong1973,Arnold1974,Gardiner2004}
\begin{equation}\label{random 200-500}
\mathbf{\xi}(t) =\frac{d\mathbf{Q}(t)}{dt},
\end{equation}
where $\mathbf{Q}(t)$ is a two-dimensional Wiener process with a diffusion constant $D_{1}$.

The mathematically rigorous form of Eq.(\ref{Langevin 300-200}) is the following stochastic differential equations based on It\^{o} stochastic integral \cite{Kallenberg1997}
\begin{equation}\label{Langevin 200-1000}
\left\{
\begin{array}{ll}
d\mathbf{r}(t) = \mathbf{v}(t)dt, \\
m_{q}d\mathbf{v}(t) = -\eta_{0}m_{d}\mathbf{v}(t)dt + \mathbf{F}(\mathbf{r}, t)dt + d\mathbf{Q}(t),
\end{array}
\right.
\end{equation}
where $\mathbf{r}(t)$, $\mathbf{v}(t)$, $\mathbf{F}(\mathbf{r}, t)$  and $\mathbf{Q}(t), t \geq 0$ are stochastic processes on a probability space $(\Omega, \mathcal{F}, P)$, $d\mathbf{Q}(t)$ are independent of all of the $\mathbf{r}(s)$, $\mathbf{v}(s)$, with $s \leq t$, $\mathbf{r}(0)=\mathbf{r}_0$, $\mathbf{v}(0)=\mathbf{v}_0$.

The microstate of the CP at time $t$ is defined by the random vector $(\mathbf{r}(t), \mathbf{v}(t))$ \cite{WangXS2014}.
\begin{proposition}\label{convergence 200-1100}
Suppose that Eqs.(\ref{variance 200-300}) are valid and the force field $\mathbf{F}(\mathbf{r}, t): R^{2} \times R_{+} \rightarrow R^{2}$
 satisfy a global Lipschitz condition, that is, for some constant $C_{0}$,
\begin{equation}\label{Lipschitz 200-1200}
|\mathbf{F}(\mathbf{r}_{1}, t) - \mathbf{F}(\mathbf{r}_{2}, t)|
\leq C_{0}|\mathbf{r}_{1} - \mathbf{r}_{2}|,
\end{equation}
for all $\mathbf{r}_{1}$ and $\mathbf{r}_{2}$ in $R^{2}$, where $R^{2}$ denotes the two-dimensional Descartes space, $R_{+}$ denotes the set of  positive real numbers. Then, at a time scale of an observer very large compare to the relaxation time $\tau_{c}$,
\begin{equation}\label{relaxation 200-1300}
\tau_{c} \equiv \frac{m_{q}}{\eta_{0}m_{d}},
\end{equation}
the solution $\mathbf{r}(t)$ of the Langevin equation Eq.(\ref{Langevin 200-1000}) converges to the solution $\mathbf{y}(t)$ of the following Smoluchowski equation Eq.(\ref{Smoluchowski 200-1500}) with probability one uniformly for t in compact subintervals of $[0, \infty)$ for all $\mathbf{v}_0$, i.e.,
\begin{equation}\label{convergence 200-1400}
\lim_{1/\tau_{c} \rightarrow \infty} \mathbf{r}(t) = \mathbf{y}(t),
\end{equation}
where $\mathbf{y}(t)$ is the solution of the following Smoluchowski equation
\begin{equation}\label{Smoluchowski 200-1500}
d\mathbf{y}(t)= \mathbf{b}(\mathbf{y}, t)dt + d\mathbf{w}(t),
\end{equation}
where $\mathbf{y}(0)=\mathbf{r}_0$, $\mathbf{w}(t)$ is a two-dimensional Wiener process with a diffusion constant $D_{3}$ defined by
\begin{equation}\label{diffusion 200-1600}
D_{3} = \frac{ k_{0} T_{\omega}}{\eta_{0}m_{d}}.
\end{equation}
\end{proposition}

A proof of Proposition \ref{convergence 200-1100} can be found in the Appendix A. Following similar methods in Ref. \cite{WangXS2014}, a Schr\"{o}dinger like equation (\ref{Schrodinger 30-10}) and Eq.(\ref{nelson 400-1500}) can be derived, refers to Appendix B.

Putting Eq.(\ref{nelson 400-1500}) into Eq.(\ref{Schrodinger 30-10}), we have the following result.
\begin{proposition}\label{reduce Schrodinger 30-20}
The Schr\"{o}dinger like equation Eq.(\ref{Schrodinger 30-10}) reduces to the following Schr\"{o}dinger like equation
\begin{equation}\label{Schrodinger 700-300}
i \hbar \frac{\partial \psi}{\partial t} =
-\frac{\hbar^2}{2m_{w}}\nabla^2\psi + U(\mathbf{r})\psi,
\end{equation}
where $m_{w}$ is the wave mass defined by Eq.(\ref{wave 400-1400}).
\end{proposition}

From Eq.(\ref{damping 300-250}), in vacuum the damping mass $m_{d}$ reduces to the mass $m_{c}$ of a CP. In vacuum, $T_{\omega}$ reduces to $T_{0}$. Therefore, in vacuum the wave mass $m_{w}$ defined by Eq.(\ref{wave 400-1400}) reduces to $m_{c}$. Thus, the Schr\"{o}dinger like equation (\ref{Schrodinger 700-300}) in the condensed matter is a generalization of the Schr\"{o}dinger equation (\ref{Schrodinger 100-100}) in vacuum.

\section{Calculation of direct current (DC) electrical conductivity \label{sec 1050}}
If there is no external magnetic field and the external electric field $\mathbf{E}$ is a constant vector field, then the Langevin equation (\ref{Langevin 300-200}) can be written as
\begin{equation}\label{Langevin 1050-100}
m_{q}\frac{d^2\mathbf{r}}{dt^2} = - \eta_{0}m_{d}\mathbf{v} + e_{c}\mathbf{E} +\mathbf{\xi}(t),
\end{equation}
where $e_{c}$ is the electric charge of the CP.

If the mean velocity $\mathbf{v}$ of the CP is high enough such that $d\mathbf{v}/dt=0$, then we call this velocity as drift velocity and denotes it as $\mathbf{v}_{d}$. Thus, if the observer look at the CP for a time long enough comparing to the relaxation time $\tau_{c}$, then he will observe the long time averaged quantities of the Langevin equation (\ref{Langevin 1050-100}). Since $(d\mathbf{v}/dt)|_{\mathbf{v}=\mathbf{v}_{d}}=0$ and $E\mathbf{\xi}(t)=0$, the long time averaged form of the Langevin equation (\ref{Langevin 1050-100}) can be written as (\cite{AshcroftNW1976}, p.\ 7; \cite{YanSS2011}, p.\ 16)
\begin{equation}\label{drift 1050-200}
\mathbf{v}_{d}=\frac{e_{c}\mathbf{E}}{\eta_{0}m_{d}}.
\end{equation}

The current density $\mathbf{j}$ corresponding to the drift velocity $\mathbf{v}_{d}$ is (\cite{AshcroftNW1976}, p.\ 7; \cite{YanSS2011}, p.\ 16)
\begin{equation}\label{current 1050-300}
\mathbf{j}=n_{c}e_{c}\mathbf{v}_{d},
\end{equation}
where $n_{c}$ is the number density of CPs.

Putting Eq.(\ref{drift 1050-200}) and Eq.(\ref{current 1050-300}), we have
\begin{equation}\label{current 1050-400}
\mathbf{j}=\frac{n_{c}e_{c}^{2}}{\eta_{0}m_{d}}\mathbf{E}.
\end{equation}

Eq.(\ref{current 1050-300}) can be written as (\cite{AshcroftNW1976}, p.\ 7)
\begin{equation}\label{current 1000-450}
j_{i}=\sum_{j}\sigma_{ij}E_{j},
\end{equation}
where $j_{i}$ is the $i$th component of the current density $\mathbf{j}$, $E_{j}$ is the $j$th component of the electric field $\mathbf{E}$, $\sigma_{ij}$ is the conductivity tensor which can be written as
\begin{equation}\label{conductivity 1000-900}
\sigma_{ij}=\sigma\delta_{ij},
\end{equation}
where
\begin{equation}\label{conductivity 1000-1000}
\sigma =\frac{n_{c}e_{c}^{2}}{\eta_{0}m_{d}},
\end{equation}
$\delta_{ij}$ is the Kronecker symbol

It is known that the resistivity of the normal states of cuprate superconductors exhibits strong anisotropy (\cite{HanRS2014}, p.\ 190). Thus, Eq.(\ref{conductivity 1000-900}) may be only valid for the plane conductivity $\rho_{ab}$ of two dimensional cuprate superconductors and not valid for bulk cuprate. Noticing $\rho_{ab}=1/\sigma$ and Eq.(\ref{conductivity 1000-1000}), we have
\begin{equation}\label{resistivity 1050-500}
\rho_{ab} = \frac{\eta_{0}m_{d}}{n_{c}e_{c}^{2}}.
\end{equation}

Using Eq.(\ref{relaxation 200-1300}), Eq.(\ref{resistivity 1050-500}) can also be written as
\begin{equation}\label{resistivity 1050-600}
\rho_{ab} = \frac{m_{q}}{n_{c}e_{c}^{2}\tau_{c}}.
\end{equation}

\section{Linear temperature dependence of resistivity in the normal states of cuprate superconductors \label{sec 1100}}
Using Eq.(\ref{damping 300-250}), Eq.(\ref{resistivity 1050-500}) can be written as
\begin{equation}\label{resistivity 1100-500}
\rho_{ab} = \frac{(\eta_{0}+\eta_{1})m_{c}}{n_{c}e_{c}^{2}}.
\end{equation}

Inspired by Eq.(\ref{resistivity 1100-500}), we speculate that the origin of the linear-in-temperature resistivity of the strange metals may be the linear temperature dependence of the damping coefficient $\eta_{0}+\eta_{1}$. Thus, we introduce the following assumption.
\begin{assumption}\label{assumption 1100-100}
Suppose that the following relationship is valid in the strange metal states of two dimensional cuprate superconductors
\begin{equation}\label{damping 1100-600}
\frac{\eta_{0}+\eta_{1}}{\eta_{0}} = b_{0}\frac{T}{T_{0}}.
\end{equation}
where $b_{0}$ ia a parameter to be determined.
\end{assumption}

In vacuum we have $T=T_{0}$ and $\eta_{1}=0$ \cite{WangXS2014}. Suppose that Eq.(\ref{damping 1100-600}) is also valid in vacuum. Thus, we have $b_{0} = 1$. Using Eq.(\ref{damping 1100-600}), Eq.(\ref{resistivity 1100-500}) can be written as
\begin{equation}\label{resistivity 1100-700}
\rho_{ab} = A_{1}T,
\end{equation}
where
\begin{equation}\label{coefficient 1100-800}
A_{1} = \frac{\eta_{0}m_{c}}{n_{c}e_{c}^{2}T_{0}}.
\end{equation}

Noticing Eq.(\ref{planck 50-60}), Eq.(\ref{resistivity 1100-700}) can also be written as
\begin{equation}\label{resistivity 1100-900}
\rho_{ab} = A_{2}T,
\end{equation}
where
\begin{equation}\label{coefficient 1100-1000}
A_{2} = \frac{2k_{0}m_{c}}{\hbar n_{c}e_{c}^{2}}.
\end{equation}

If $A_{2}$ is independent of temperature $T$, then Eq.(\ref{resistivity 1100-700}) shows that the plane resistivity $\rho_{ab}$ is a linear function of temperature $T$. Thus, Eq.(\ref{resistivity 100-100}) is derived based on the stochastic quantization model of CP in two dimensional cuprate superconductors.

A prediction of Eq.(\ref{resistivity 1100-900}) is that if $T=0$, then $\rho_{ab}=0$, i.e., the residual resistivity $\rho_{0}=0$.

\section{Linear temperature dependence of scattering rate of CPs \label{sec 1200}}
Using Eq.(\ref{damping 300-250}) and Eq.(\ref{inertial 300-350}), Eq.(\ref{relaxation 200-1300}) can be written as
\begin{equation}\label{scatter 1200-400}
\frac{1}{\tau_{c}} = \frac{\eta_{0}+\eta_{1}}{1+\eta_{2}}.
\end{equation}

According to Eq.(\ref{scatter 1200-400}), the origin of the linear-in-temperature scattering rate $1/\tau_{c}$ of the cuprate strange metals may be the linear temperature dependence of the damping coefficient $\eta_{0}+\eta_{1}$. Noticing Eq.(\ref{damping 1100-600}), Eq.(\ref{scatter 1200-400}) can be written as
\begin{equation}\label{scatter 1200-500}
\frac{1}{\tau_{c}} = c_{1}T,
\end{equation}
where
\begin{equation}\label{coefficient 1200-600}
c_{1} = \frac{\eta_{0}}{T_{0}(1+\eta_{2})}.
\end{equation}

Noticing Eq.(\ref{planck 50-60}), Eq.(\ref{scatter 1200-500}) can also be written as
\begin{equation}\label{scatter 1200-700}
\frac{1}{\tau_{c}} = c_{2}\frac{k_{B}T}{\hbar},
\end{equation}
where
\begin{equation}\label{coefficient 1200-800}
c_{2} = \frac{2k_{0}}{k_{B}(1+\eta_{2})}.
\end{equation}

If $c_{1}$ is independent of temperature $T$, then Eq.(\ref{scatter 1200-500}) shows that the scattering rate $1/\tau_{c}$ is a linear function of temperature $T$.

If we suppose that $c_{2} \approx 1$, then Eq.(\ref{scatter 100-400}) is derived. Thus, we may say that CPs in the cuprate strange metals are undertaking the Planckian dissipation \cite{TaupinM2022}.

\section{Conclusion \label{sec 4000}}
The origin of the linear-in-temperature resistivity of the normal state of hole-doped cuprate superconductors is a unsolved problem. Inspired by the Drude formula of resistivity, we speculate that the transport scattering rate of CPs in the normal states of cuprate superconductors may be linear-in-temperature. Thus, a clue to explain the linear-in-temperature scaling law of resistivity in strange metal states of cuprate superconductors is to seek a Drude like formula of resistivity and investigate the relaxation time of CP dynamics. We suppose that a CP in a condensed matter will experiences not only a damping force exerted by vacuum but also an additional damping force exerted by the condensed matter. Thus, a Langevin equation of a CP in two dimensional condensed matter is established. Following a similar method of Nelson's stochastic mechanics, generalized Schr\"{o}dinger equation in condensed matter is derived. If CPs move with a constant velocity, then the electrical current density corresponding to the drift velocity can be calculated. Therefore, a Drude like formula of resistivity of CPs is derived. The damping coefficient of CPs in two dimensional cuprate superconductors is supposed to be a linear function of temperature. Thus, the plane resistivity and scattering rate of CPs turn out to be also linear functions of temperature.

\section*{Appendix A: Proof of Proposition \ref{convergence 200-1100}}\label{sec 5000}
We introduce the following definitions
\begin{eqnarray}
\beta &=& \frac{\eta_{0}m_{d}}{m_{q}},\label{definition 2000-110}\\
\mathbf{K}(\mathbf{r}, t) &=& \frac{\mathbf{F}(\mathbf{r}, t)}{m_{q}},\label{definition 2000-120}\\
\mathbf{B}(t) &=& \frac{\mathbf{Q}(t)}{m_{q}}.\label{definition 2000-130}
\end{eqnarray}

Then, $\mathbf{B}(t)$ is a two-dimensional Wiener process with a diffusion constant $D_{2}$ \cite{Kallenberg1997}
\begin{equation}\label{diffusion 2000-200}
D_{2} =  \frac{D_{1}}{m_{q}^2} = \frac{\eta_{0}m_{d} k_{0}T_{\omega}}{m_{q}^2} = \frac{\beta k_{0}T_{\omega}}{m_{q}}.
\end{equation}

Using Eqs.(\ref{definition 2000-110}-\ref{definition 2000-130}), Eq.(\ref{Langevin 200-1000}) can be written as
\begin{equation}\label{Langevin 2000-300}
\left\{
\begin{array}{ll}
d\mathbf{r}(t) = \mathbf{v}(t)dt, \\
d\mathbf{v}(t)
 = -\beta \mathbf{v}(t)dt
 +\mathbf{K}(\mathbf{r}, t)dt
 + d\mathbf{B}(t).
\end{array}
\right.
\end{equation}

We introduce the following definitions
\begin{eqnarray}
\mathbf{b}(\mathbf{r}, t) &=& \frac{\mathbf{K}(\mathbf{r}, t)}{\beta},\label{definition 2000-410}\\
\mathbf{w}(t) &=& \frac{\mathbf{B}(t)}{\beta}.\label{definition 2000-420}
\end{eqnarray}

Noticing Eq.(\ref{definition 2000-110}), $\mathbf{w}(t)$ is a two-dimensional Wiener process with a diffusion constant $D_{3}$ \cite{Kallenberg1997}
\begin{equation}\label{diffusion 2000-500}
D_{3} =  \frac{D_{2}}{\beta^2} = \frac{ k_{0} T_{\omega}}{\eta_{0}m_{d}}.
\end{equation}

Using Eq.(\ref{definition 2000-410}-\ref{definition 2000-420}), Eq.(\ref{Langevin 2000-300}) can be written as
\begin{equation}\label{Langevin 2000-600}
\left\{
\begin{array}{ll}
d\mathbf{r}(t) = \mathbf{v}(t)dt, \\
d\mathbf{v}(t) = -\beta \mathbf{v}(t)dt + \beta \mathbf{b}(\mathbf{r}, t)dt + \beta d\mathbf{w}(t).
\end{array}
\right.
\end{equation}

Let $\mathbf{r}(t)$ be the solution of Eq.(\ref{Langevin 2000-600}) with $\mathbf{r}(0)=\mathbf{r}_0, \mathbf{v}(0)=\mathbf{v}_0$.
According to Eq.(\ref{Lipschitz 200-1200}), the functions $\mathbf{b}(\mathbf{r}, t): R^{2} \times R_{+} \rightarrow R^{2}$ also satisfies a global Lipschitz condition. For a time scale of an observer very large compare to the relaxation time $\tau_{c} \equiv 1/\beta$, he concludes that $\beta$ can be regarded as infinity, i.e., $\beta \rightarrow +\infty$. Applying Nelson's Theorem 10.1 (\cite{NelsonE2001}, p.\ 59), the solution $\mathbf{r}(t)$ of the Langevin equation Eq.(\ref{Langevin 2000-600}) converges to the solution $\mathbf{y}(t)$ of the Smoluchowski equation Eq.(\ref{Smoluchowski 200-1500}) with probability one uniformly for $t$ in compact subintervals of $[0, \infty)$ for all $\mathbf{v}_0$. $\square$

\section*{Appendix B: Generalized Schr\"{o}dinger equation of a CP in two dimensional condensed matter}\label{sec 6000}
Noticing the asymmetry in time $t$, we can introduce the following Langevin equation \cite{Nelson1966}
\begin{equation}\label{Langevin 400-100}
\left\{
\begin{array}{ll}
d\mathbf{r}(t) = \mathbf{v}(t)dt, \quad \mathbf{r}(0)=\mathbf{r}_0\\
md\mathbf{v}(t) = -f\mathbf{v}(t)dt + \mathbf{F}(\mathbf{r}, t)dt + d\mathbf{Q}_{*}(t),
\end{array}
\right.
\end{equation}
where $d\mathbf{N}_{*}(t)$ are independent of all of the $\mathbf{r}(s)$,
$\mathbf{v}(s)$, with $s \geq t$, $\mathbf{v}(0)=\mathbf{v}_0$.

We define the following mean forward derivative $D\mathbf{y}(t)$ and the mean backward derivative $D_{*}\mathbf{y}(t)$ \cite{Nelson1966}
\begin{equation}\label{forward 2-10}
 D\mathbf{y}(t) = \lim_{\triangle t\rightarrow 0+}
 E_{t}\left [\frac{\mathbf{y}(t + \triangle t)-\mathbf{y}(t)}
 {\triangle t}\right ],
\end{equation}
\begin{equation}\label{backward 2-10}
 D_{*}\mathbf{y}(t) = \lim_{\triangle t\rightarrow 0+}E_{t} \left [ \frac{\mathbf{y}(t)-\mathbf{y}(t-\triangle t)}{\triangle
 t}\right ],
\end{equation}
where $E_{t}$ denotes the conditional expectation given the state of the system at time $t$.

We also have another Smoluchowski equation \cite{Nelson1966}
\begin{equation}\label{Smoluchowski 300-600}
 d\mathbf{y}(t) = \mathbf{b}_{*}(\mathbf{y}, t)dt + d\mathbf{w}_{*}(t),
\end{equation}
where $\mathbf{w}_{*}(t)$ has the same properties
as $\mathbf{w}(t)$ except that the
$d\mathbf{w}_{*}(t)$ are independent of the
$\mathbf{y}(s)$ with $s\geq t$.

Based on Eq.(\ref{forward 2-10}-\ref{backward 2-10}), we have \cite{Nelson1966}
\begin{equation}\label{forward 300-710}
 D\mathbf{y}(t) = \mathbf{b}(\mathbf{y}, t),
\end{equation}
\begin{equation}\label{backward 300-720}
\quad D_{*}\mathbf{y}(t) = \mathbf{b}_{*}(\mathbf{y}, t).
\end{equation}

For the probability density $\rho(\mathbf{y},t)$ of $\mathbf{y}$, we have the following forward Fokker-Planck equation and the backward Fokker-Planck equation \cite{Nelson1966}
\begin{equation}\label{Fokker 300-810}
\frac{\partial \rho}{\partial t} = - \nabla \cdot (\rho \mathbf{b}) + D_{3} \nabla^{2} \rho,
\end{equation}
\begin{equation}\label{Fokker 300-820}
\frac{\partial \rho}{\partial t} = - \nabla \cdot (\rho \mathbf{b}_{*}) - D_{3} \nabla^{2} \rho,
\end{equation}
where $\nabla \cdot \equiv \partial / \partial r_{1} + \partial / \partial r_{2}$
is the divergence operator in the two dimensional Cartesian coordinate $\{r_{1}, r_{2}\}$, $\nabla^2 \equiv \partial^2/\partial r_{1}^2 +\partial^2/\partial r_{2}^2$ is the two dimensional Laplace operator.

We introduce the definitions of current velocity $\mathbf{v}_{1}(t)$ and osmotic velocity $\mathbf{u}_{1}(t)$ \cite{Nelson1966}
\begin{equation}\label{velocity 20-11}
 \mathbf{v}_{1} = \frac{1}{2}(\mathbf{b}+\mathbf{b}_{*}),
\end{equation}
\begin{equation}\label{velocity 20-12}
\quad \mathbf{u}_{1} = \frac{1}{2}(\mathbf{b}-\mathbf{b}_{*}).
\end{equation}

The current velocity $\mathbf{v}_{1}(t)$ is the deterministic part of the total velocity $\mathbf{b}(t)$ of the CP. The osmotic velocity $\mathbf{u}_{1}(t)$ is the stochastic part of the total velocity $\mathbf{b}(t)$. The non-zero osmotic velocity $\mathbf{u}_{1}(t)$ is a difference between stochastic mechanics deterministic mechanics \cite{WangXS2014}.

We have the following result \cite{Nelson1966}:
\begin{equation}\label{osmotic velocity 20-20}
\mathbf{u}_{1} = D_{3} \frac{\nabla \rho}{\rho} = D_{3} \nabla(\ln\rho).
\end{equation}

We introduce the following definition of wave mass.
\begin{definition}\label{microstate 3-10}
The wave mass of the particle is defined by
\begin{equation}\label{wave 400-1400}
m_{w} \equiv \frac{T_{0}}{T_{\omega}}m_{d}.
\end{equation}
\end{definition}

Using Eq.(\ref{planck 50-60}), Eq.(\ref{wave 400-1400}) and Eq.(\ref{diffusion 200-1600}), we have
\begin{equation}\label{nelson 400-1500}
D_{3} = \frac{\hbar}{2m_{w}}.
\end{equation}

Similar to the method of Ref. \cite{Nelson1966}, we introduce the following definition of osmotic potential $R_{1}$
\begin{equation}\label{osmotic velocity 80-420}
m_{w}\mathbf{u}_{1} = \nabla R_{1},
\end{equation}
where the osmotic potential $R_{1}$ is defined by
\begin{equation}\label{R1 80-430}
R_{1} \triangleq m_{w} D_{3} \ln \rho.
\end{equation}

We introduce the definition of the mean second derivative $\mathbf{a}(t)$ of the stochastic process $\mathbf{y}(t)$ \cite{Nelson1966}
\begin{equation}\label{mean second derivative 2-10}
 \mathbf{a}(t) = \frac{1}{2}DD_{*}\mathbf{y}(t) +  \frac{1}{2}D_{*}D\mathbf{y}(t).
\end{equation}

Applying a similar method of E. Nelson \cite{Nelson1966}, we can derive the following Proposition \ref{velocity vu 40-10} \cite{WangXS2014}.
\begin{proposition}\label{velocity vu 40-10}
The current velocity field $\mathbf{v}_{1}(\mathbf{r},t)$ and the osmotic velocity field $\mathbf{u}_{1}(\mathbf{r},t)$ satisfy the following coupled equations:
\begin{eqnarray}
\frac{\partial \mathbf{v}_{1}}{\partial t}&=&\frac{\mathbf{F}}{m_{w}}-(\mathbf{v}_{1}\cdot\nabla)\mathbf{v}_{1}+(\mathbf{u}_{1}\cdot\nabla)\mathbf{u}_{1}\nonumber \\
&&+D_{3}\nabla^2\mathbf{u}_{1}, \label{Nelson equation 20-21}\\
\frac{\partial \mathbf{u}_{1}}{\partial t}&=&-D_{3}\nabla(\nabla\cdot\mathbf{v}_{1})-\nabla(\mathbf{v}_{1}\cdot\mathbf{u}_{1}). \label{Nelson equation 20-22}
\end{eqnarray}
\end{proposition}

Similar to the deterministic newtonian mechanics, we can also introduce the following concept of deterministic momentum field $\mathbf{p}_{d}(\mathbf{r},t)$ and  stochastic momentum field $\mathbf{p}_{s}(\mathbf{r},t)$ of the Brownian particle:
\begin{equation}\label{momentum 20-11}
\mathbf{p}_{d}(\mathbf{r},t) = m_{w}\mathbf{v}_{1}(\mathbf{r},t),
\end{equation}
\begin{equation}\label{momentum 20-12}
\mathbf{p}_{s}(\mathbf{r},t) = m_{w}\mathbf{u}_{1}(\mathbf{r},t).
\end{equation}

\begin{proposition}\label{momentum pdps 40-10}
If there exists a functions $S_{1}(\mathbf{r},t)$ such that
\begin{equation}\label{assumption grad 20-20}
\mathbf{p}_{d} = \nabla S_{1},
\end{equation}
then, the deterministic momentum field $\mathbf{p}_{d}(\mathbf{r},t)$ and  stochastic momentum field $\mathbf{p}_{s}(\mathbf{r},t)$ of the Brownian particle satisfy the following equations
\begin{eqnarray}
 \frac{\partial \mathbf{p}_{d}(t)}{\partial t} &=&
\mathbf{F} - \frac{1}{2m_{w}} \nabla (\mathbf{p}_{d}^{2}) +
\frac{1}{2m_{w}} \nabla (\mathbf{p}_{s}^{2})\nonumber \\
&& + D_{3} \nabla^2 \mathbf{p}_{s}, \label{momentum pd 20-11}\\
\frac{\partial \mathbf{p}_{s}(t)}{\partial t} &=& - D_{3}
\nabla^2 \mathbf{p}_{d}
- \frac{1}{m_{w}}\nabla(\mathbf{p}_{d}
\cdot \mathbf{p}_{s}). \label{momentum ps 20-12}
\end{eqnarray}
\end{proposition}

{\bfseries{Proof of Proposition \ref{momentum pdps 40-10}}}.
We have the following equations in field theory:
\begin{equation}\label{formula field 20-10}
\nabla (\mathbf{a}  \cdot \mathbf{b})
=  (\mathbf{a}  \cdot \nabla ) \mathbf{b}
+ (\mathbf{b}  \cdot \nabla ) \mathbf{a}
+ \mathbf{a} \times ( \nabla \times \mathbf{b})
+ \mathbf{b} \times ( \nabla \times \mathbf{a}),
\end{equation}
\begin{equation}\label{formula field 20-20}
\nabla^2 \mathbf{a}
=  \nabla(\nabla \cdot \mathbf{a}) - \nabla \times ( \nabla \times \mathbf{a}),
\end{equation}
\begin{equation}\label{formula field 20-30}
\nabla \times (\nabla \varphi) = 0,
\end{equation}
where $\mathbf{a}$ and $\mathbf{b}$ are arbitrary vectors, $\varphi$ is an arbitrary scalar function.

Using Eq.(\ref{osmotic velocity 80-420}), Eq.(\ref{assumption grad 20-20}), Eq.(\ref{formula field 20-10}) and Eq.(\ref{formula field 20-30}), we have
\begin{equation}\label{formula v 20-11}
\frac{1}{2}\nabla (\mathbf{v}_{1}^{2}) = (\mathbf{v}_{1}
\cdot \nabla)\mathbf{v}_{1},
\end{equation}
\begin{equation}\label{formula u 20-12}
\frac{1}{2}\nabla (\mathbf{u}_{1}^{2}) = (\mathbf{u}_{1}
\cdot \nabla)\mathbf{u}_{1}.
\end{equation}

Using Eq.(\ref{assumption grad 20-20}), Eq.(\ref{formula field 20-20}) and Eq.(\ref{formula field 20-30}), we have
\begin{equation}\label{formula v 20-20}
\nabla^2 \mathbf{v}_{1} = \nabla(\nabla \cdot \mathbf{v}_{1}).
\end{equation}

Putting Eq.(\ref{formula v 20-11}-\ref{formula u 20-12}) into Eq.(\ref{Nelson equation 20-21}) and using Eq.(\ref{osmotic velocity 80-420}) and  Eq.(\ref{assumption grad 20-20}), we obtain Eq.(\ref{momentum pd 20-11}).
Putting Eq.(\ref{formula v 20-20}) into Eq.(\ref{Nelson equation 20-22}) and using Eq.(\ref{osmotic velocity 80-420}) and  Eq.(\ref{assumption grad 20-20}), we obtain Eq.(\ref{momentum ps 20-12}). $\square$

We may call the functions $S_{1}(\mathbf{r},t)$ defined in Eq.(\ref{assumption grad 20-20}) as the current potential. The current potential $S_{1}(\mathbf{r},t)$ is not uniquely defined by the deterministic momentum field $\mathbf{p}_{d}(\mathbf{r},t)$. For instance, let $S_{1}^{'} = S_{1} + c_{0}$, where $c_{0}$ is an arbitrary constant. Then, we also have $\nabla (S_{1}^{'}) = \mathbf{p}_{d}$.

\begin{theorem}\label{generalized Hamilton-Jacobi 40-10}
If there exist two functions $U(\mathbf{r})$ and $S_{1}$ such that
\begin{equation}\label{assumption grad 20-15}
\mathbf{F}(\mathbf{r},t)= - \nabla U(\mathbf{r}),
\end{equation}
\begin{equation}\label{assumption grad 20-25}
\mathbf{p}_{d} = \nabla S_{1},
\end{equation}
then, the generalized Hamilton's principal function
\begin{equation}\label{principal 400-3000}
W_{1} \triangleq S_{1} -i R_{1}
\end{equation}
satisfies the following generalized Hamilton-Jacobi equation
\begin{eqnarray}
-\frac{\partial W_{1}}{\partial t} &=& \frac{1}{2m_{w}} (\nabla W_{1})^2 + U(\mathbf{r}) \nonumber \\
&& - i D_{3}\nabla^{2} W_{1}  + \theta_{1}(t) + i \theta_{2}(t),\label{ghj 20-10}
\end{eqnarray}
where $\theta_{1}(t)$ and $\theta_{2}(t)$ are two unknown real functions of $t$.
\end{theorem}

{\bfseries{Proof of Theorem \ref{generalized Hamilton-Jacobi 40-10}.}} We multiply Eq.(\ref{momentum pd 20-11}) with $-1$
and then plus Eq.(\ref{momentum ps 20-12}) multiplied by $i$. Thus, we obtain
\begin{eqnarray}
-\frac{\partial (\mathbf{p}_{d} -i \mathbf{p}_{s})}{\partial t} &=& - \mathbf{F} + \frac{1}{2m_{w}} \nabla [(\mathbf{p}_{d}-i \mathbf{p}_{s})^2]\nonumber \\
&&- i D_{3}\nabla^{2} (\mathbf{p}_{d} -i \mathbf{p}_{s}). \label{ghj 20-20}
\end{eqnarray}

We introduce the following definition
\begin{equation}\label{momentum p 20-10}
\mathbf{p} \triangleq \mathbf{p}_{d}
-i \mathbf{p}_{s}, \quad  i^{2} = -1.
\end{equation}

Thus, Eq.(\ref{ghj 20-20}) becomes
\begin{equation}\label{ghj 20-30}
-\frac{\partial \mathbf{p}}{\partial t}
= - \mathbf{F} + \frac{1}{2m_{w}} \nabla (\mathbf{p}^2) - i D_{3}\nabla^{2} \mathbf{p}.
\end{equation}

We may regard the function $S_{1}$ and $R_{1}$
as the deterministic part and stochastic part of a generalized Hamilton's principal function $S$ defined by
\begin{equation}\label{Hamilton function s 20-10}
W_{1} \triangleq S_{1} -i R_{1}, \quad  i^{2} = -1.
\end{equation}

Putting Eq.(\ref{osmotic velocity 80-420}) and  Eq.(\ref{assumption grad 20-25}) into Eq.(\ref{momentum p 20-10}) and using Eq.(\ref{Hamilton function s 20-10}), we have
\begin{equation}\label{nabla 20-10}
\mathbf{p} = \nabla W_{1}.
\end{equation}

Putting Eq.(\ref{nabla 20-10}) into Eq.(\ref{ghj 20-30}), we obtain
\begin{equation}\label{ghj 20-40}
-\frac{\partial (\nabla W_{1})}{\partial t} = - \mathbf{F} + \frac{1}{2m_{w}} \nabla [(\nabla W_{1})^2]   - i D_{3}\nabla^{2} (\nabla W_{1}).
\end{equation}

Noticing $\mathbf{F} = - \nabla U(\mathbf{r})$, Eq.(\ref{ghj 20-40}) becomes
\begin{equation}\label{ghj 20-50}
-\frac{\partial \nabla (W_{1})}{\partial t}
= \nabla V + \frac{1}{2m_{w}} \nabla[(\nabla W_{1})^2]   - i D_{3}\nabla^{2} (\nabla W_{1}).
\end{equation}

Eq.(\ref{ghj 20-50}) can be written as
\begin{equation}\label{ghj 20-60}
\nabla \left [\frac{\partial W_{1}}{\partial t}
+  U(\mathbf{r}) + \frac{1}{2m_{w}} \nabla W_{1}^2   - i D_{3}\nabla^{2} (\nabla W_{1})\right ] = 0.
\end{equation}

Integration of Eq.(\ref{ghj 20-60}) gives
\begin{eqnarray}
- \frac{\partial W_{1}}{\partial t} &=& U(\mathbf{r}) + \frac{1}{2m_{w}} \nabla W_{1}^2 - i D_{3}\nabla^{2} (\nabla W_{1}) \nonumber \\
&&+ \theta_{1}(t) + i \theta_{2}(t), \label{ghj 20-70}
\end{eqnarray}
where $\theta_{1}(t)$ and  $\theta_{2}(t)$ are two unknown real functions of $t$. $\square$

The generalized Hamilton's principal function $W_{1} \triangleq S_{1} -i R_{1}$ is not uniquely defined by $\mathbf{p}_{d}$. The reason is that $\mathbf{p}_{d} = \nabla S_{1}$. Thus, $S_{1}$ is not uniquely defined by $\mathbf{p}_{d}$.

Similar to Bohr's correspondence principle, we may also introduce the following correspondence principle in stochastic mechanics.
\begin{assumption}\label{correspondence principle 20-10}
If the diffusion constant $D_{3}$ is small enough, i.e., $D_{3} \rightarrow 0$, then, the generalized Hamilton-Jacobi equation Eq.(\ref{ghj 20-10}) in stochastic mechanics becomes identical to the following Hamilton-Jacobi equation in classical mechanics \cite{Goldstein2002}
\begin{equation}\label{Hamilton-Jacobi 200-100}
-\frac{\partial W}{\partial t} = \frac{1}{2m} (\nabla W)^2  + U(\mathbf{r}),
\end{equation}
where $W(\mathbf{r},t)$ is a real function called Hamilton's principal function, $U(\mathbf{r})$ is a potential.
\end{assumption}

\begin{theorem}\label{theorem ghj 20-20}
Suppose that the assumptions Eq.(\ref{assumption grad 20-15} - \ref{assumption grad 20-25}) are valid.
Then, the generalized Hamilton's principal function $W_{1}$
satisfies the following generalized Hamilton-Jacobi equation
\begin{equation}\label{generalized Hamilton-Jacobi 20-50}
-\frac{\partial W_{1}}{\partial t} = \frac{1}{2m_{w}} (\nabla W_{1})^2
+ U(\mathbf{r}) - i D_{3}\nabla^{2} W_{1}.
\end{equation}
\end{theorem}

{\bfseries{Proof of Theorem \ref{theorem ghj 20-20}}}
Let $D_{3} = 0$. Then, from Eq.(\ref{osmotic velocity 20-20}), we have $\mathbf{u}_{1} = 0$. Thus, from Eq.(\ref{momentum 20-12}), we have $\mathbf{p}_{s} = 0$. Then, from Eq.\ref{osmotic velocity 80-420},  $R_{1}$ is a constant. Thus, Eq.(\ref{ghj 20-10}) can be written as
\begin{equation}\label{hj 20-10}
-\frac{\partial S_{1}}{\partial t} = \frac{1}{2m_{w}} (\nabla S_{1})^2 + U(\mathbf{r}) + \theta_{1}(t) + i \theta_{2}(t),
\end{equation}

According to Assumption \ref{correspondence principle 20-10}, Eq.(\ref{hj 20-10}) should be identical to the Hamilton-Jacobi equation Eq.(\ref{Hamilton-Jacobi 200-100}). Thus, we obtain $\theta_{1}(t) = 0$ and $\theta_{2}(t) = 0$. $\square$

Similar to the Hamiltonian mechanics \cite{Goldstein2002}, we introduce the following definition of wave function
\begin{equation}\label{wave function 30-10}
\psi(\mathbf{r},t) = \exp{\left[ \frac{i W_{1}(\mathbf{r},t)}{2m_{w}D_{3}}\right ]}.
\end{equation}

The generalized Hamilton's principal function $W_{1}$ is not uniquely defined by the deterministic momentum field $\mathbf{p}_{d}$. Therefore, the wave function $\psi(\mathbf{r},t)$ defined by Eq.(\ref{wave function 30-10}) is not uniquely defined by $\mathbf{p}_{d}$.

\begin{theorem}\label{Schrodinger like 30-10}
The wave function $\psi(\mathbf{r},t)$ satisfies the following Schr\"{o}dinger like equation
\begin{equation}\label{Schrodinger 30-10}
i  \frac{\partial \psi}{\partial t} = - D_{3} \nabla^2\psi
+ \frac{1}{2m_{w}D_{3}} U(\mathbf{r})\psi.
\end{equation}
Eq.(\ref{Schrodinger 30-10}) is equivalent to the generalized Hamilton-Jacobi equation Eq.(\ref{generalized Hamilton-Jacobi 20-50}).
\end{theorem}

{\bfseries{Proof of Theorem \ref{Schrodinger like 30-10}}}.
From the definition Eq.(\ref{wave function 30-10}), we have
\begin{equation}\label{Hamilton function 30-10}
W_{1}(\mathbf{r},t) = \frac{2m_{w}D_{3}}{i} \ln \psi(\mathbf{r},t).
\end{equation}

Putting Eq.(\ref{Hamilton function 30-10}) into Eq.(\ref{generalized Hamilton-Jacobi 20-50}),
we obtain a Schr\"{o}dinger like equation Eq.(\ref{Schrodinger 30-10}).
Conversely, putting Eq.(\ref{wave function 30-10})
into Eq.(\ref{Schrodinger 30-10}), we obtain the generalized Hamilton-Jacobi equation Eq.(\ref{generalized Hamilton-Jacobi 20-50}). $\square$

\providecommand{\noopsort}[1]{}\providecommand{\singleletter}[1]{#1}%

\end{document}